\documentclass[pre, amsmath, onecolumn, showpacs, superscriptaddress]{revtex4}
\usepackage{graphicx}
\usepackage{mathtools}
\usepackage{amsfonts}
\usepackage{bbold}
\usepackage{bm}
\usepackage{multirow}
\usepackage{xcolor}
\usepackage{tikz}
\usetikzlibrary{shapes,shapes.multipart}

\begin{document}

\title{Winding statistics of a Brownian particle on a ring}
\author{Anupam Kundu}
\affiliation{Laboratoire de Physique Th\'eorique et Mod\`eles Statistiques (UMR 8626 du CNRS),
Universit\'e Paris-Sud, B\^at.\ 100, 91405 Orsay Cedex, France}
\author{Alain Comtet}
\affiliation{Laboratoire de Physique Th\'eorique et Mod\`eles Statistiques (UMR 8626 du CNRS),
Universit\'e Paris-Sud, B\^at.\ 100, 91405 Orsay Cedex, France}
\affiliation{Universit\'e Pierre et Marie Curie Paris 6, 75005 Paris, France}
\author{Satya N. Majumdar}
\affiliation{Laboratoire de Physique Th\'eorique et Mod\`eles Statistiques (UMR 8626 du CNRS),
Universit\'e Paris-Sud, B\^at.\ 100, 91405 Orsay Cedex, France}

\begin{abstract}
We consider a Brownian particle moving on a ring. We study the probability distributions of the total number of turns 
and the net number of counter-clockwise turns the particle makes till time $t$. Using a method based on the renewal properties of Brownian walker, 
we find exact analytical expressions of these distributions. This method serves as an alternative to the standard path integral techniques which are not always 
easily adaptable for certain observables. For large $t$, we show that these distributions have Gaussian scaling forms. We also compute 
large deviation functions associated to these distributions characterizing atypically large fluctuations. 
We provide numerical simulations in support of our analytical results.
\end{abstract}

\maketitle

\section{Introduction}
\noindent 
Starting from the pioneering works of Edwards \cite{Edwards67,Edwards68}, statistical studies of winding properties of topologically constrained random processes 
have been a subject of keen interest in various contexts such as in the physics of polymers 
\cite{Rudnik87,Rudnik88, Grosberg03, wiegel}, the fluxlines in superconductors \cite{Nelson88, Drossel96}
and many others. The winding properties of planar Brownian paths have also been a subject of interest to mathematicians for a long time.
In 1958, Spitzer \cite{Spitzer58} 
studied the distribution of the total angle $\theta(t)$ wound by a planar 
Brownian path around a prescribed point in time $t$. He showed that in the large time 
limit the scaled random variable $\frac{2\theta(t)}{\ln t}$ has a Cauchy distribution \emph{i.e.} with infinite first moment. 
Later, various generalizations and extensions of this classic result have been put forward \cite{Rudnik87,Rudnik88}. Further asymptotic laws of planar 
Brownian motion unifying and extending this approach to the case of $n$ different points were obtained by Pitman and Yor \cite{Pittman86}. Recently, 
winding properties of other planar processes such as time correlated Gaussian process \cite{Doussal09},  
Schramm-Loewner evolution\cite{Schram}, loop erased random walk \cite{Hagendrof08} and conformally invariant curves \cite{Duplantier02} have also been studied.

\begin{figure}[h]
\includegraphics[scale=0.4]{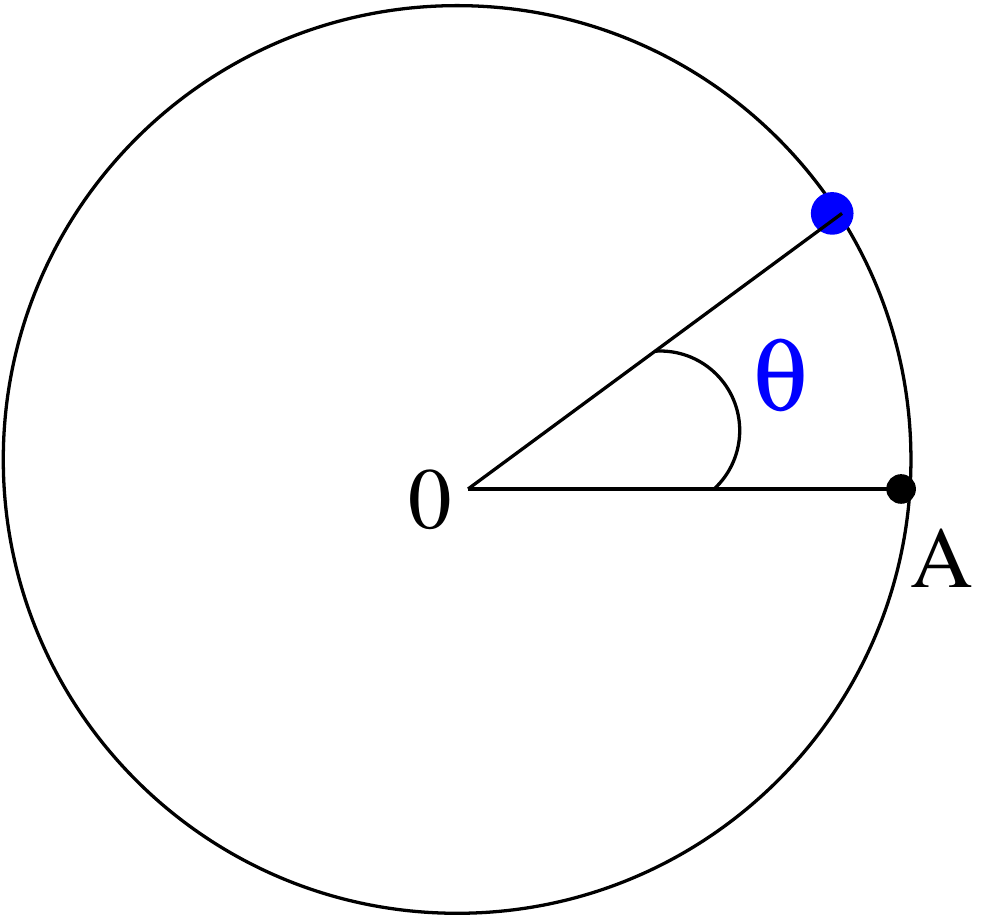}
\caption{(Color online) (a) A Brownian particle diffusing on a ring of unit radius. }
\label{fig1}
\end{figure}

For constrained Brownian trajectories on a surface, the methods of computing probabilities associated with the winding properties are mathematically similar to 
the path integral of a quantum particle coupled to magnetic fields \cite{Brereton87, Khandekar88, Comtet90, Comtet91}. This similarity has been exploited in various 
situations, for example, the algebraic number of full turns a polymer 
makes around a cylinder on which it lays, can be analyzed by studying the physics of a charged quantum particle interacting with a tube of magnetic 
flux \cite{Nelson97}. 
Thermal fluctuation in the trajectory of a vortex defect in the superconducting order parameter \cite{Nelson89}, which are characterized by winding numbers, 
can be analyzed by studying path integrals of a polymer melt. These kind of path integral techniques have also been used to study other quantities, 
like e.g. the area between a Brownian path and its subtending chord \cite{Duplantier89, Comtet90, Comtet91}. Other functionals of the Brownian motion 
that occur in the context of weak localization have been discussed in \cite{Comtet05}.

In this paper we study winding properties of a Brownian particle moving on a ring of unit radius (see Fig. \ref{fig1}) analytically.
In mesoscopic physics, such diffusion on a ring geometry appears important in quantum transport through a connected ring. 
The weak localization correction to the classical conductance of the ring can be computed from the knowledge of the    
net number of complete turns around the ring \cite{Texier}. 
Here we are, in particular, interested in the distributions of total number of turns $n(t)$ and the net number of   
counter-clockwise turns $k(t)$ the particle makes around the circle till time $t$ in the following two different cases: 
(a) {\it Free}: the position $\theta(t)$ of the particle at time $t$ is not constrained and, (b) {\it Constrained} :
the position $\theta(t)$ of the particle at time $t$ is constrained to be the 
starting point (A in Fig. \ref{fig1}) \emph{i.e.} $\theta(t)=2\pi l$ where $l$ is 
an integer. The path configurations in case (b) are called Brownian bridges on a circle. 
If $\theta(t)$ represents the position of the particle on the circle, then its stochastic evolution is given by the following equations
\begin{eqnarray}
\frac{d \theta}{dt}= \sqrt{2D}~\xi(t),~~
\langle \xi(t) \rangle = 0,~~~~~\langle \xi(t)\xi(t') \rangle = \delta(t-t'), \label{langevin}
\end{eqnarray}
where $D$ is the diffusion constant and $\xi(t)$ is a mean zero white Gaussian noise. As time increases, the particle turns around the circle 
both clockwise and counter-clockwise. Let $n_+(t)$ be the number of full counter clockwise turns and $n_-(t)$ be the number of full clockwise turns 
the particle makes in time $t$, then the total number of full turns and the net number of full counter-clockwise 
turns are given by $n(t)=n_+(t) + n_-(t)$ and $k(t)=n_+(t) - n_-(t) $ respectively. Clearly, $n(t)$ 
and $k(t)$ are random variables and they are called the total winding number and net winding number respectively. We would like to compute the distributions 
$P(n,t)$ of $n(t)$, $R(k,t)$ of $k(t)$ and also the joint distribution $P(n,k,t)$ for both the {\it{free}} and the {\it{constrained}} cases. In 
previous studies for example in the context of polymers \cite{wiegel}, in the context planar Brownian trajectories \cite{Comtet90} 
or in the context of vortex lines in type II superconductors \cite{Nelson97}, path integral 
techniques have been successfully used to evaluate winding statistics of net winding numbers, where one maps the problem of finding the probabilities 
to the problem of finding the propagator of a suitable quantum particle. 
In contrast, here it seems adapting the path integral techniques is not easy. For example, it is hard to 
implement the path integral techniques for computing the distributions of quantities like the total winding number $n(t)$ and the net winding number $k(t)$ 
especially in the {\it{free}} case. The main purpose of this paper is to present an alternative method based on {\it{renewal}} 
properties of the process that allows us to compute the winding distribution exactly.

Let us present a brief summary of our results. 
We study the statistics of the total winding number and the net winding number of single Brownian particle on the ring for the {\it free} 
and the {\it constrained} case separately. First, we consider the {\it free} case. For this case, we obtain exact expressions for 
the mean and variance of both the total winding number $n$ and the net winding number $k$ as a function of time $t$. We see that for large $t$, the 
mean total winding number grows linearly as $\langle n(t) \rangle \sim \frac{D}{2\pi^2}t$ and its variance also grows linearly as 
$V(t) = \langle n(t)^2 \rangle - \langle n(t) \rangle^2 \simeq \frac{D}{3\pi^2}t$. On the other hand, the mean net winding number 
$\langle k(t) \rangle=0$. In fact all odd order moments of $k$ are zero. This is because the probabilities of getting net winding number $k$ and 
$-k$ at any time $t$, are equal when the particle starts from the origin. The variance of $k$ grows as 
$\langle k(t)^2 \rangle|_{t \to \infty}  \simeq \frac{D}{2\pi^2}t$  which can be understood simply by 
writing the position of the walker $\theta(t)$ as 
$\theta(t)=2\pi k(t) + \zeta(t)$ where $\zeta(t)$ corresponds to the displacement of the walker in the time
 remaining after the last complete turn. Since the displacement 
$\zeta(t)$ is bounded in $(-2\pi~:~2\pi)$, we have $\langle k^2(t)\rangle \simeq \langle 
\left (\frac{ x(t) }{2\pi^2} \right)^2 \rangle=\frac{D}{2\pi^2}t$ for large $t$.
These large $t$ behavior of the moments suggests, that the random variable $\chi_{tot} = \frac{n-\frac{Dt}{2\pi^2}}{\sqrt{t}}$, 
constructed from the random variable $n$, will have a $t$ independent distribution in the large $t$ limit. Similarly, the random variable 
$\chi_{net} = \frac{k}{\sqrt{t}}$ will also have a $t$ independent distribution in the large $t$ limit. This means, the 
distribution $P(n,t)$ of the total winding number $n$ and the distribution $R(k,t)$ of the net winding number $k$, have the following scaling forms 
$P(n,t) \simeq \frac{1}{\sqrt{t}}G\left(\frac{n-\frac{Dt}{2\pi^2}}{\sqrt{t}}\right)$ and 
$R(k,t) \simeq \frac{1}{\sqrt{t}}H\left(\frac{k}{\sqrt{t}}\right)$
respectively. We compute the scaling functions exactly and they are given by simple Gaussians, 
\begin{equation}
G(y)= \sqrt{\frac{3\pi}{2D}}\text{exp}\left[ -\frac{3\pi^2}{2D} y^2\right],~~~~\text{and}~~~~
H(y)= \sqrt{\frac{\pi}{D}}\text{exp}\left[ -\frac{\pi^2}{D} y^2\right].\label{Gauss}
\end{equation}
We have verified these scaling distributions numerically. Here we mention that, the above scaling distributions are valid for large $t$ 
and over the region where the fluctuations of $n$ and $k$ around their respective means are typical \emph{i.e.}  
$\lesssim \mathcal{O}(\sqrt{t})$. When these fluctuations around their respective means are larger than 
$\mathcal{O}(\sqrt{t})$, then the probabilities can not be described by the scaling distributions in Eq. (\ref{Gauss}). The probabilities of 
atypically large fluctuations of $\mathcal{O}(t)$ are described by large deviation functions. 
We will see in Sec. (\ref{LDFs}) that, for large $n$ and $t$ but $\frac{n}{t}$ fixed, and similarly, for large $k$ and $t$ but $\frac{k}{t}$ fixed,
the distributions $P(n,t)$ and $R(k,t)$ have the following large deviation forms  
\begin{eqnarray}
 P(n,t) &&\overset{n \to \infty,~t \to \infty}{\underset{\frac{n}{t}~\text{fixed}}{\xRightarrow{\hspace*{1.5cm}}}}~
~\text{exp}\left[ -t~\mathcal{G}\left(\frac{n}{t}\right)\right], \label{LDF-tot} \\
R(k,t) &&\overset{k \to \infty,~t \to \infty}{\underset{\frac{k}{t}~\text{fixed}}{\xRightarrow{\hspace*{1.5cm}}}}~
~\text{exp}\left[ -t~\mathcal{H}\left(\frac{k}{t}\right)\right].\label{LDF-net}
\end{eqnarray}
In this paper, we compute these large deviation functions $\mathcal{G}(x)$ and $\mathcal{H}(x)$. 
We also find exact analytical expressions of $P(n,t)$, $R(k,t)$ and the joint distribution $P(n,k,t)$ for arbitrary $n$ and $k$ at any time $t$.

Next we study the case where the trajectories of the Brownian particle are constrained to be exactly at $2\pi l$ with $l =0,\pm1,\pm2,\pm3...$ at time $t$. 
In this case  we denote the probability distributions of $n$ and $k$ by $P_c(n,t)$ and $R_c(k,t)$ respectively and the joint distribution by $P_c(n,k,t)$ where 
the subscript ``c'' stands for {\it constrained} trajectories. We find exact expressions of these distributions at any time $t$. 
Moreover, we show that in the large $t$ limit, the scaling forms and the large deviation functions associated to the distributions $P_c(n,t)$ and $R_c(k,t)$ are 
exactly same as in the previous case \emph{i.e} the scaling distributions are described by the same functions $G(y)$ and $H(y)$ whereas the large deviation functions 
are also described by $\mathcal{G}(x)$ and $\mathcal{H}(x)$.

The paper is organized as follows. In section \ref{free} we study the winding statistics for the {\it{free}} case 
\emph{i.e.} for unconstrained Brownian motion on the ring. In the beginning of this section we discuss in detail 
about the connection between making a complete turn around the 
circle and making first exit from a box of size $4\pi$. Then in the subsection \ref{MVS-dis} we compute the time dependence of the 
moments of $n$ and $k$ and also their scaling distributions. In the next subsection \ref{LDFs} we find the large deviation forms associated to the distributions $P(n,t)$ 
and $R(k,t)$. In subsection \ref{exact} we derive exact explicit expressions of the $P(n,t)$ and $R(k,t)$ for arbitrary $n$ and $k$. After completing the analysis for the 
{\it{free}} case, in section \ref{Brow-brdg} we study the moments and the distributions $P_c(n,t)$ and $R_c(k,t)$ corresponding to $n$ and $k$ for Brownian bridge on a 
ring. In the last section \ref{conclusion} we present our conclusion. Few details are left in the Appendix.

\section{Un-constrained Brownian motion on the ring}
\label{free}
\begin{figure}[h]
\includegraphics[scale=0.4]{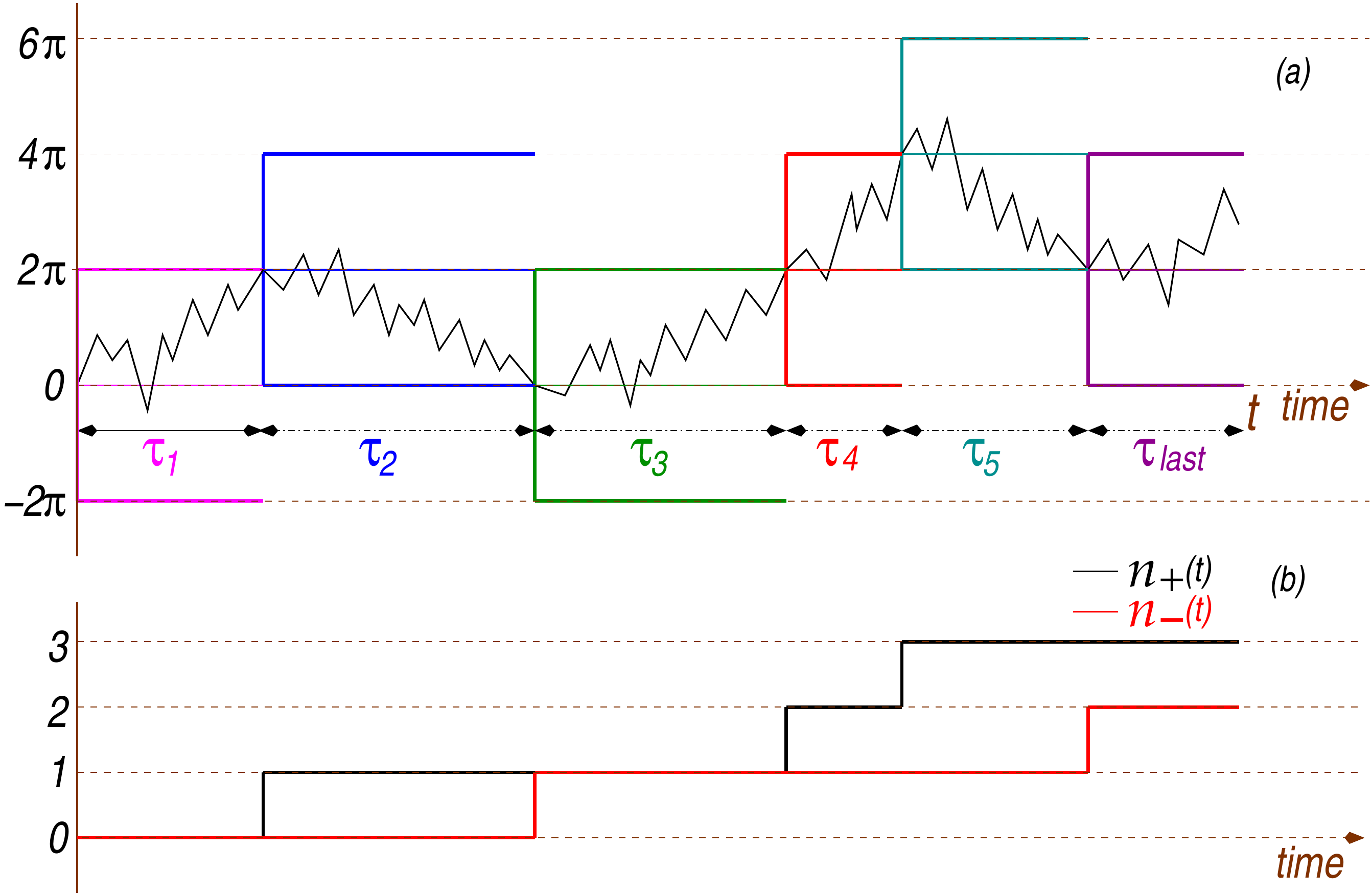}
\caption{(Color online)  The top panel of this plot contains a typical trajectory (black zig-zag line) of the particle over time $t$. 
This particular trajectory wounds the circle around 
completely five times in time intervals $\tau_1,\tau_2,...,\tau_5$ successively. Each of these complete turn around the circle is associated to the first exit of 
the particle from a box of size $4\pi$ given that the particle started from the center of the box. After each complete turn this first exit process gets renewed with 
the center of the box shifted to the particle's current position as shown by the different colored boxes. The bottom panel shows the increase 
of $n_+$ and $n_-$ with time corresponding to this particular trajectory. }
\label{fig1a}
\end{figure}

\noindent
As mentioned in the introduction, the renewal properties of the Brownian walker helps us to find the distributions of 
the number of complete counter-clockwise turns $n_+(t)$ and the number of complete 
clockwise turns $n_-(t)$ around the circle in time $t$. To see how, let us look at (see Fig. \ref{fig1a}) 
a typical trajectory of the Brownian particle on the ring, starting at the origin. 
In Fig. \ref{fig1a} we see that the particle makes its first full turn around the circle at time $\tau_1$ when it makes, 
starting at the origin, a first exit from the box $[-2\pi,~2\pi]$ (magenta box).
In the next time interval $[\tau_1, \tau_1+\tau_2]$ the particle, starting at $2\pi$, 
makes a first exit from the box $[0,~4\pi]$ (blue box) and thus completing its second full 
turn around the circle at time $\tau_1+\tau_2$. Similar first exit events happen for each of the  
successive complete turns at later time intervals except in the last time interval where the particle may not have enough time to perform a complete turn. 
So we see that, each complete turn around the circle is associated to a first exit from a box of size $4\pi$ given that the particle starts from 
the center of the box and after each complete turn the location of the center of the box gets shifted to the particle's current position to perform similar first exit 
for the next complete turn \emph{i.e.} to renew the first exit process for the next complete turn. 
The first exit through the upper 
boundary of the box corresponds to a full counter-clockwise turn around the circle whereas the first exit through the lower boundary corresponds 
to a full clockwise turn around 
the circle. Using this connection with first exit problem from a box of length $4\pi$ and the renewal property of a Brownian walker we study the distributions of having 
total winding number $n$ and net winding number $k$ in time $t$ for both {\it free} and {\it constrained} cases.

\begin{figure}[h]
\includegraphics[scale=0.3]{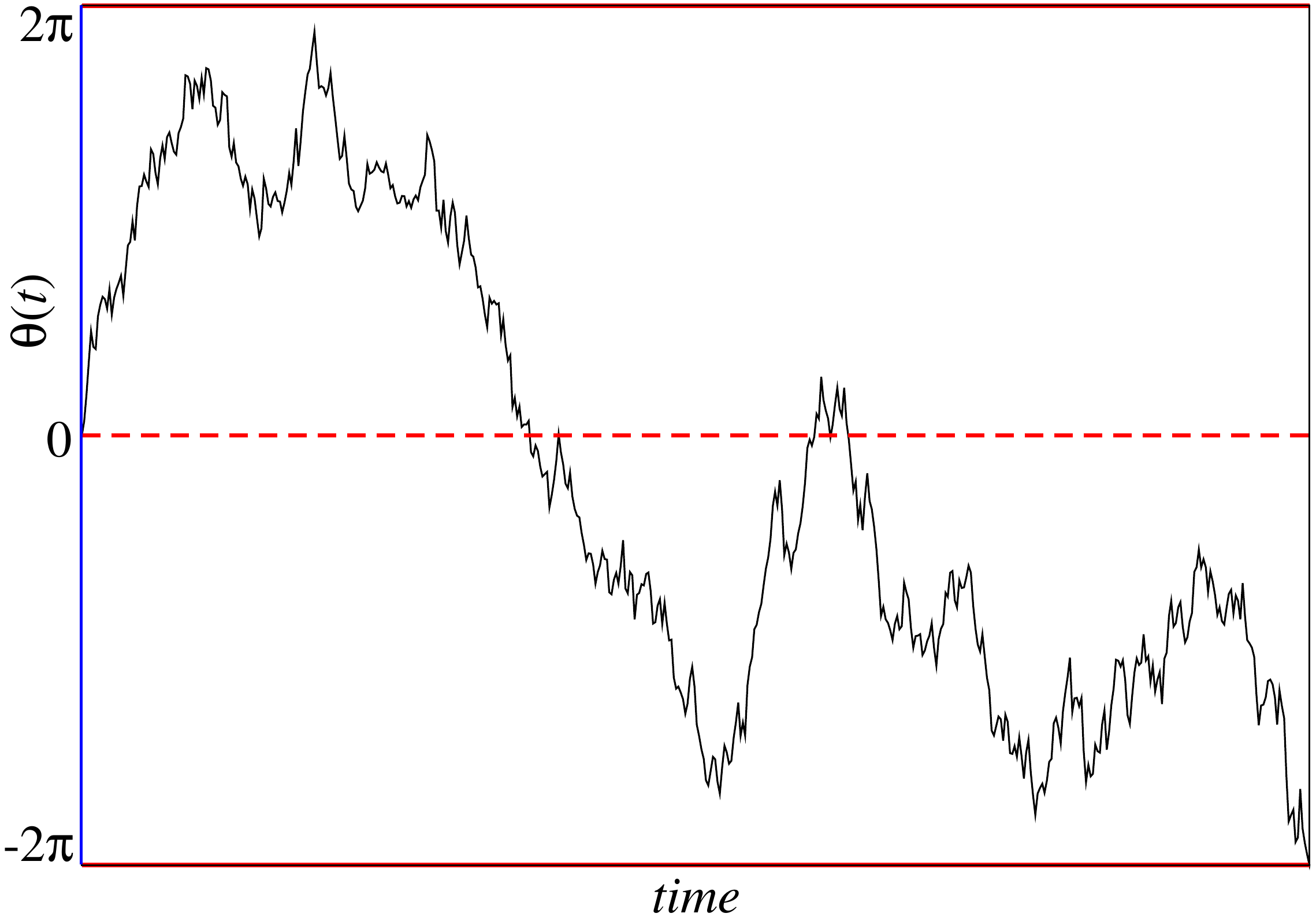}
\caption{(Color online) First exit of the Brownian particle through the boundaries of the box $[-2\pi,~2\pi]$.}
\label{fig2}
\end{figure}

If $f(t)$ represents the first exit probability distribution of the particle from the box of size $4\pi$ given that it had 
started from the center of the 
box, then the probability that the particle starting from the origin, 
makes a complete counter-clockwise turn in time $t$ to $t+dt$, 
is given by $\frac{f(t)}{2}dt$.
Similarly, by symmetry (as the particle starts at the center of the box), the probability that it makes 
a clockwise turn in time $t$ to $t+dt$, 
is also given by $\frac{f(t)}{2}dt$.
To compute $f(t)$, it turns out to be convenient to study 
$q(x,t)$ which represents the survival probability that the particle, 
starting from position $\theta(0)=x$, stays inside the box $[-2\pi,~2\pi]$ (see Fig. \ref{fig2}) 
till time $t$. Knowing $q(x,t)$, the first exit probability is then given by 
\begin{equation}
f(t)= - \frac{d q(t)}{d t}~~\text{where}~~~q(t)\equiv q(0,t). \label{FPndSVl}
\end{equation}
The survival probability $q(x,t)$ satisfies the following Backward Fokker-Planck equation 
\cite{RednerBook,Satya05,Bray13}
\begin{equation}
\frac{\partial q(x,t) }{\partial t} = D \frac{\partial^2 q(x,t)}{\partial x^2},\label{BFP1} 
\end{equation}
with initial condition $q(x,0)=1$ and boundary conditions (BCs) $q(x=\pm 2\pi,t)=0$. To solve the above equation it is convenient to use the Laplace 
transform (LT) and inverse Laplace transform (ILT) which for a general function $g(t)$ are defined as follows: 
\begin{eqnarray}
\text{LT~of}~g(t): &&\tilde{g}(s) = \mathcal{L}_s[g(t)] = \int_0^{\infty}dt~ e^{-st}g(t)~, \nonumber \\
\text{ILT~of}~\tilde{g}(s):&& g(t) =\mathcal{L}_t^{-1}[\tilde{g}(s)] =\frac{1}{2\pi i} \int_{\mathcal{B}} ds ~\tilde{g}(s)~e^{st}, \label{LT-def}
\end{eqnarray}
where $\mathcal{B}$ stands for the Bromwich integral in the complex $s$ plane.
Taking the Laplace transform $\tilde{q}(x,s) = \mathcal{L}_s\left[q(x,t) \right]$ 
on both sides of Eq. (\ref{BFP1}), we get $D \frac{\partial^2 \tilde{q}(x,s)}{\partial x^2} - s\tilde{q}(x,s)=-1$
with BCs $\tilde{q}(x=\pm 2\pi,s)=0$, whose solution at $x=0$ is given by 
\begin{equation}
 \tilde{q}(s)\equiv \tilde{q}(0,s) = \frac{1}{s} \left[1-\text{sech} \left(2\pi\sqrt{\frac{s}{D}}\right) \right],~~
 \text{which~provides}~~q(t)=\mathcal{L}_t^{-1}[\tilde{q}(s)]. \label{qtilde}
\end{equation}
Now taking derivative of $q(t)$ with respect to $t$ we get $f(t)=-dq(t)/dt$ from Eq. (\ref{FPndSVl}) whose LT is given by 
\begin{equation}
 \tilde{f}(s) = \mathcal{L}_s[f(t)]=1-s\tilde{q}(s) = \text{sech} \left(2\pi\sqrt{\frac{s}{D}}\right). \label{ftilde}
\end{equation}

Knowing $q(t)$ and $f(t)$, we proceed to compute $P(n,t)$ as follows. Consider the event of $n$ complete turns in $[0,t]$ and let
$\{\tau_1,\tau_2,\ldots, \tau_n\}$ denote their respective durations (see Fig. \ref{fig1a}). The difference
$\tau_{\rm last}= t- \sum_{i=1}^n {\tau_i}$ denotes the duration of the last unfinished turn. 
The probability of such an event, where both the number of turns $n$ as well as their durations $\{\tau_i\}'s$  
are random variables, is denoted by $\mathcal{P}(n,~\{\tau_1,\tau_2,....,\tau_n,\tau_{last}\};~t)$
and can be expressed as
\begin{equation}
\mathcal{P}(n,~\{\tau_1,\tau_2,....,\tau_n,\tau_{last}\};~t) = f(\tau_1)f(\tau_2)...f(\tau_n)~q(\tau_{\rm last})~
\delta \left( t-\tau_{\rm last}-\sum \limits_{i=1}^n \tau_i\right) \label{JPDF1}
\end{equation}
where $q(\tau)$ is the survival probability in Eq. (\ref{qtilde}) 
and $f(\tau)=-dq(\tau)/d\tau$ is the first exit probability from the box. 
The first $n$ factors involving $f$ in Eq. (\ref{JPDF1}) represent $n$ complete turns (or the first exits
from the box of length $4\pi$ as explained in Fig. \ref{fig1a}), the last factor $q(\tau_{\rm last})$ represents
the unfinished turn (\emph{i.e.}, the probability to stay inside the box during $\tau_{\rm last}$ as shown
in Fig. \ref{fig1a}). Finally the delta function represents the fact that the durations of all these intervals
add up to $t$. In writing this joint distribution in Eq. (\ref{JPDF1}), we have used the renewal
property of the Brownian motion, \emph{i.e.}, the successive intervals are statistically independent.

Similar renewal equations appeared before in the study of the record statistics of discrete time 
random walkers \cite{MajumZiff,wergen,Schehr14, Godreche14}. In the context of record statistics, $q(t)$ represents 
the probability that the walker 
stays below its starting position up to $t$ steps and $f(t)=q(t-1)-q(t)$ 
represents the first-passage probability that the walker crosses 
its starting position from below for the first time in between steps $t-1$ and $t$.
In contrast here we have continuous time $t$ and $q(t)$ and $f(t)$ represent respectively the
survival and the first exit probability from a finite box $[0,4\pi]$.

The probability $P(n,t)$ that the particle makes $n$ complete turns either clockwise or anti clockwise in time $t$,  
can then be obtained from the joint probability density in Eq. (\ref{JPDF1}) by integrating over
all the durations
\begin{eqnarray}
 P(n,t) = \int_0^{t}d\tau_1\int_0^{t}d\tau_2...\int_0^{t}d\tau_n \int_0^{t}d\tau_{\rm last} 
 ~~\mathcal{P}(n,~\{\tau_1,\tau_2,....,\tau_n,\tau_{last}\};~t). \label{P_n-1}
\end{eqnarray}
As we will see later, instead of working with these multiple time integrals, it is simpler to work in Laplace space. 
Using the convolution structure in Eq. (\ref{P_n-1}), we perform Laplace transform with respect to $t$ and get  
\begin{eqnarray}
 \tilde{P}(n,s)&=&\mathcal{L}_s[P(n,t)]=\int_0^{\infty}e^{-st}p(n,t)~dt, \nonumber \\
 &=&\left[\tilde{f}(s) \right]^n\tilde{q}(s)= \frac{1}{s}~\frac{\cosh\left(2\pi\sqrt{\frac{s}{D}}\right)-1}
 {\left[\cosh\left(2\pi\sqrt{\frac{s}{D}}\right)\right]^{n+1}}, \label{LTP_n}
\end{eqnarray}
where we have used Eq. (\ref{qtilde}) and (\ref{ftilde}). 
From the above expression one can easily check that, $\sum_{n=0}^{\infty}P(n,t)=1$ as 
$\sum_{n=0}^{\infty}\tilde{P}(n,s)=s^{-1}$.
Finally, taking ILT of $\tilde{P}(n,s)$ given explicitly above, we get the probability $P(n,t)$ that the particle has total winding number $n$ in time $t$ as 
\begin{equation}
 P(n,t) = \frac{1}{2\pi i} \int_{\mathcal{B}} ds~e^{st}~\left[\tilde{f}(s) \right]^n\tilde{q}(s). \label{ILTP_n}
\end{equation}

Once we know $P(n,t)$, the joint probability $P(n,k,t)$ that the particle has total winding number $n$ and net winding number $k$ in time $t$,  
can simply be obtained as follows: Given the particle makes a complete turn around the circle at time $t$ to $t+dt$ with probability $f(t)dt$, the 
probability that it makes a counter-clockwise turn is $f_+(t)dt=\frac{f(t)}{2}dt$ and similarly, the 
probability that it makes a clockwise turn is also $f_-(t)dt=\frac{f(t)}{2}dt$.
This is because the probabilities associated to the first exits trough the upper and lower boundaries are equal when the particle starts from the center of the box.
Hence, given that the particle makes a complete turn at time $t$, the probability that this turn will be a 
counter-clockwise turn is $\frac{1}{2}$ and the probability that this turn will be a clockwise turn is also $\frac{1}{2}$. So the probability that there are 
$m$ counter clockwise turns out of $n$ total complete turns is given by 
${n \choose m} \left(\frac{1}{2}\right)^m~\left(\frac{1}{2}\right)^{n-m}$ where the binomial factor represents the number of ways to choose 
$m$ counter-clockwise turns from $n$ total complete turns. If the particle has total winding number $n$ and net winding number $k$, 
there must be $n_+=\frac{n+k}{2}$ complete counter-clockwise turns and the rest $n_-=\frac{n-k}{2}$ clockwise turns in time $t$. 
Hence, the joint probability of having  total winding number $n$ and 
 net winding number $k$ in time $t$, is then given by $ P(n,k,t)=\text{Prob.}[n_+=\frac{n+k}{2}~\text{counter-clockwise~turns}~$ $\text{given~that~there~are~}
 n~\text{total~no.~of~turns}]$ $
 ~\times~\text{Prob.}\left[n~\text{total~no.~of~turns~in~time}~t\right] $ \emph{i.e.}
\begin{eqnarray}
 P(n,k,t) &=& {n \choose \frac{n+k}{2}}~\left(\frac{1}{2}\right)^n~P(n,t). \label{Joint-free}
\end{eqnarray}
From this joint probability, the marginal probability $R(k,t)$ of having net winding number $k$ in time $t$, is obtained by summing over $n$ as follows
\begin{equation}
 R(k,t) = \sum \limits_{m=0}^{\infty} P(2m+|k|,k,t),
\end{equation}
whose LT can be expressed using Eq. (\ref{ILTP_n}) as 
$\tilde{R}(k,s) = \sum \limits_{m=0}^{\infty}{2m+|k| \choose m}~\left(\frac{\tilde{f}(s)}{2}\right)^{2m+|k|}\tilde{q}(s)$. Now using the identity \cite{wiki}
\begin{equation}
 \sum \limits_{m=0}^{\infty}{2m+|k| \choose m}z^m = \frac{1}{\sqrt{1-4z}}\left(\frac{1-\sqrt{1-4z}}{2z} \right)^{|k|};~|z| < \frac{1}{4},
\end{equation}
and performing some algebraic simplification we get
\begin{equation}
\tilde{R}(k,s) = \mathcal{L}_s[R(k,t)] = \frac{1}{s} \tanh\left(\pi\sqrt{\frac{s}{D}}\right)~\text{exp}\left(-2\pi |k|\sqrt{\frac{s}{D}} \right). \label{LTR-k}
\end{equation}
 one can find explicit expressions of $P(n,t)$ and $R(k,t)$ after performing the inverse Laplace transforms of $\tilde{P}(n,s)$ and $\tilde{R}(k,s)$. 
 Before doing that, let us try to understand whether $P(n,t)$ and $R(k,t)$ have scaling distributions by looking at how the mean and 
 variance of $n$ and $k$ grow with time $t$. 

\subsection{Moments and scaling distributions }
\label{MVS-dis}
 \noindent 
The $m$th moment of the total winding number $n$ can formally be written as 
\begin{eqnarray}
 \langle n(t)^m \rangle &=& \sum \limits_{n=0}^{\infty} n^m~P(n,t) 
 =  \frac{1}{2\pi i} \int_{\mathcal{B}} ds~e^{st}~\sum \limits_{n=0}^{\infty} n^m~\tilde{P}(n,s),
\end{eqnarray}
where $\tilde{P}(n,s)$ is given explicitly in Eq. (\ref{LTP_n}). One can write a similar expression for the $m$th moment of the net winding number $k$, with only 
difference is that $\tilde{P}(n,s)$ will now be replaced by $\tilde{R}(k,s)$. Looking at the expression of $\tilde{R}(k,s)$ in Eq. (\ref{LTR-k}) closely, we see  
$\tilde{R}(k,s) = \tilde{R}(-k,s)$, which implies that all odd order moments of $k$ are zero. 
Using the expressions of $\tilde{P}(n,s)$ and $\tilde{R}(k,s)$ from Eqs. (\ref{LTP_n}) and (\ref{LTR-k}), and performing some 
algebraic manipulations, one can show that
\begin{eqnarray}
 \langle n(t) \rangle =\langle k(t)^2 \rangle &=&  \frac{1}{2}~\frac{1}{2\pi i} \int_{\mathcal{B}} dq~\frac{1}{q~\sinh^2(\sqrt{q})}e^{\frac{qDt}{\pi^2}} 
 \label{mean-contour}\\
 \langle n(t)^2 \rangle &=&\frac{1}{4}~\frac{1}{2\pi i}\int_{\mathcal{B}} dq~\frac{\cosh(2\sqrt{q})+1}{q~\sinh^4(\sqrt{q})}~e^{\frac{qDt}{\pi^2}}.
 \label{2ndmom-contour}
\end{eqnarray}
\begin{figure}[h]
\includegraphics[scale=0.3]{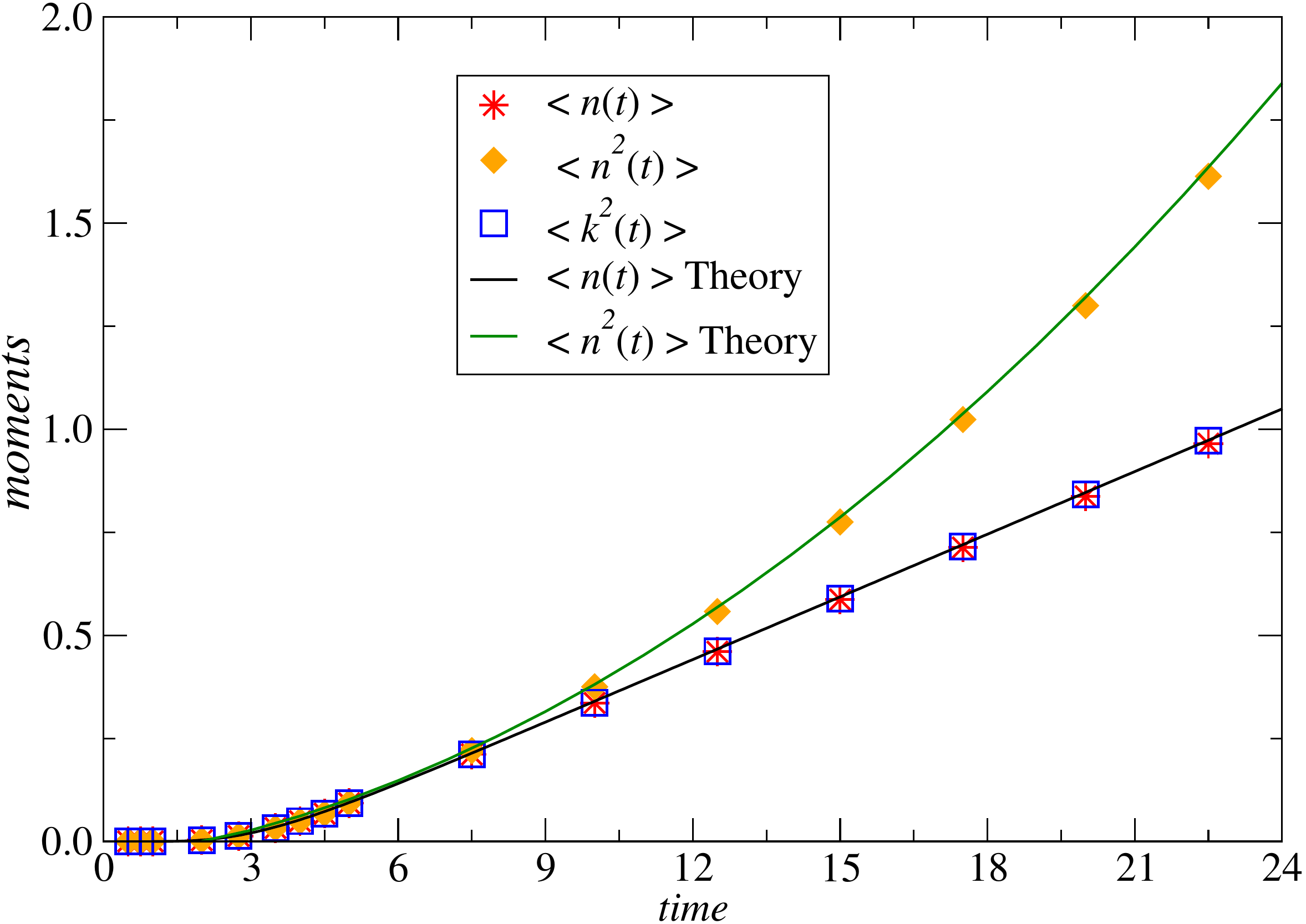}
\caption{(Color online) Comparison of the moments obtained through direct numerical simulation of the Langevin equation (\ref{langevin}) with the analytically 
obtained expressions in Eqs. (\ref{mean-appndx}) and (\ref{second-appndx}).}
\label{fig3}
\end{figure}
These Bromwich integrals can be computed by computing residues 
at the poles $q_l = -l^2\pi^2$ for $l=0, 1,2,3...$ of $\text{cosech}(\sqrt{q})$. Exact expressions of 
$\langle n(t) \rangle$, $\langle n(t)^2 \rangle$ and $\langle k(t)^2 \rangle$ are given 
in Eqs. (\ref{mean-appndx}) and (\ref{second-appndx}). In Fig. (\ref{fig3}) we compare these analytical expressions with the same obtained from direct numerical 
simulations of the Langevin equation in Eq. (\ref{langevin}) and see nice agreement.
From these analytical expressions one can easily see the following small and large $t$ behaviors 
\begin{eqnarray}
\langle n(t) \rangle \simeq 
\begin{cases}
 & \frac{2\sqrt{Dt}}{\pi\sqrt{\pi}}~\text{exp}\left(-\frac{\pi^2}{Dt} \right);~~t\to 0 \\
 & \\
 & \left(\frac{Dt}{2\pi^2}-\frac{1}{6}\right);~~~~~~~~~~~t \to \infty,
\end{cases}
~~~\text{and}~~~
\langle n(t)^2 \rangle  \simeq
\begin{cases}
 &\frac{2\sqrt{Dt}}{\pi\sqrt{\pi}}~\text{exp}\left(-\frac{\pi^2}{Dt} \right); ~~t\to 0 \\
 &\\
 & \frac{D^2t^2}{4\pi^4} + \frac{Dt}{6\pi^2} -\frac{2}{45} ;~~~~t \to \infty.
\end{cases}
\end{eqnarray}
Such small and large $t$ behavior of the mean and variance, 
can also be obtained, respectively, from the large and small $q$ behaviors of the integrands in the Eqs. (\ref{mean-contour}) and (\ref{2ndmom-contour}).
Following similar procedure one can easily show from Eq. (\ref{Joint-free}) and Eq. (\ref{ILTP_n}) that the correlation between $n$ and $k$ for large $t$, grows as 
$\sim t^{3/2}$.
\begin{figure}[h]
\includegraphics[scale=0.3]{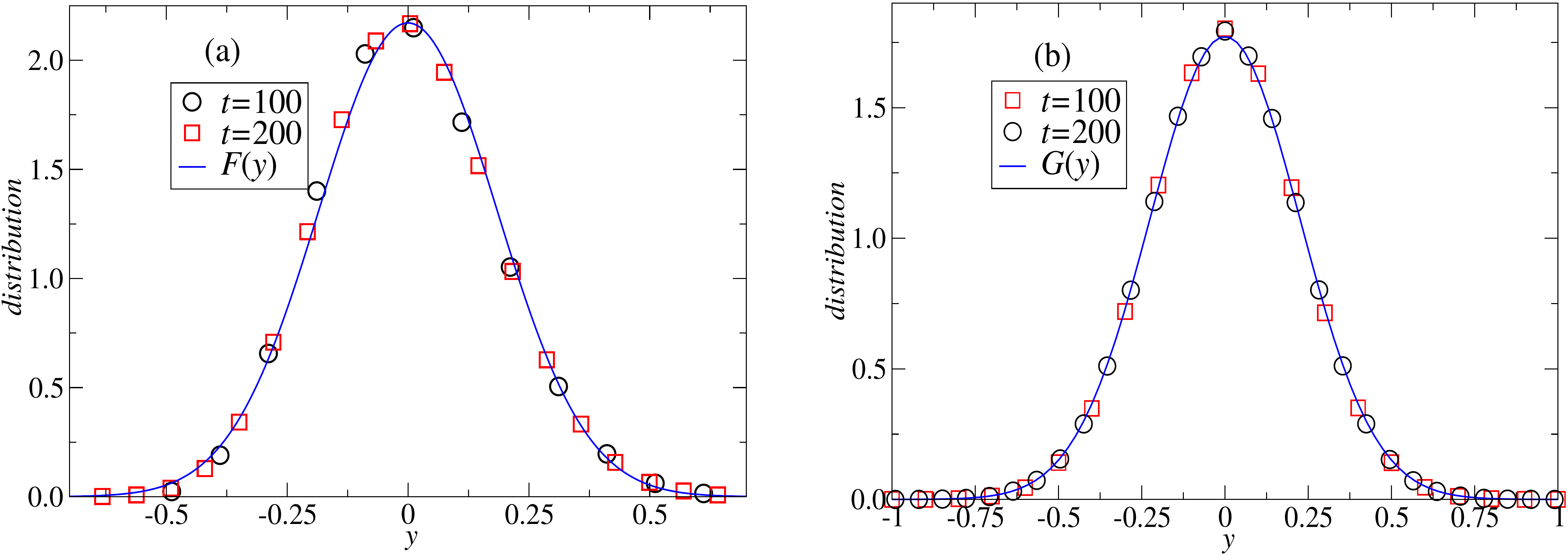}
\caption{(Color online) In figure (a) we compare the distribution $P(n,t)$ obtained from simulation with the scaling function $G(y)$ by plotting 
$\sqrt{t}P(n,t)$ vs. $(n-\frac{Dt}{2\pi^2})/\sqrt{t}$ along with $G(y)$ vs. $y$. In figure (b) we compare the distribution $R(k,t)$ 
obtained from simulation with the scaling function $H(y)$ by plotting 
$\sqrt{t}R(k,t)$ vs. $k/\sqrt{t}$ along with $H(y)$ vs. $y$}
\label{fig4}
\end{figure}

The linear growths of $\sigma^2_n(t)=\langle n(t)^2 \rangle-\langle n(t) \rangle^2$ and $\langle k(t)^2 \rangle$ at large $t$, 
suggest that if we scale the fluctuations of $n$ and $k$ around their mean by $\sqrt{t}$ then 
the distributions $P(n,t)$ and $R(k,t)$ will have the following scaling forms 
$P(n,t) \simeq \frac{1}{\sqrt{t}}G\left(\frac{n-\frac{Dt}{2\pi^2}}{\sqrt{t}}\right)$ and 
$R(k,t) \simeq \frac{1}{\sqrt{t}}H\left(\frac{k}{\sqrt{t}}\right)$ where the scaling functions $G(y)$ and $H(y)$ are given in Eq. (\ref{Gauss}). 
In Fig. (\ref{fig4}) we verify these scaling forms by comparing them with $P(n,t)$ and $R(k,t)$ 
obtained from direct simulation of the Langevin equation (\ref{langevin}). In  Fig. (\ref{fig4}a) we plot numerically obtained $\sqrt{t}~P(n,t)$ against 
$(n-\frac{Dt}{2\pi^2})/\sqrt{t}$ for $t=100$ and $t=200$ and on comparison with $G(y)$ (blue line) we see nice agreement. On the other hand, 
in Fig. (\ref{fig4}b) we plot numerically obtained $\sqrt{t}~R(k,t)$ against $\frac{k}{\sqrt{t}}$ for $t=100$ and $t=200$, to compare with the scaling distribution 
$H(y)$ (blue line). Here also we see nice agreement. The scaling forms of $P(n,t)$ and $R(k,t)$, specified by the scaling functions $G(y)$ and $H(y)$ respectively, are 
valid for large $t$ and over the regions, where their fluctuations around their respective means are $\lesssim \mathcal{O}(\sqrt{t})$. 
When these fluctuations around their respective means are of $\mathcal{O}(t)$, then the probability distributions are described by large deviation tails.

\subsection{Large deviation forms of the distributions $P(n,t)$ and $R(k,t)$}
\label{LDFs}
\noindent
In the previous subsection we have studied the scaling distributions associated to $P(n,t)$ and $R(k,t)$, which describe the typical fluctuations of $n$ and 
$k$ around their respective means. In this subsection, we will study the distributions of atypically large fluctuations which are described by the large deviation tails. 
We will see that for large $n$ and $t$ but $\frac{n}{t}$ fixed, and similarly, for large $k$ and $t$ but $\frac{k}{t}$ fixed,
the distributions $P(n,t)$ and $R(k,t)$ have large deviation forms as given in Eqs. (\ref{LDF-tot}) and (\ref{LDF-net}).
Let us first focus on deriving the large deviation form of the probability distribution $P(n,t)$ of the total winding number $n$. 
Taking $n \to \infty$ and $t \to \infty$ limit while keeping $n/t$ finite, one can write the right hand side of Eq. (\ref{ILTP_n}) 
in the following form 
\begin{eqnarray}
 && P(n,t) \simeq \frac{1}{2\pi i} \int_{\mathcal{B}} ds~~e^{t[\Psi(s,x) + \mathcal{O}(1/t)]}, \nonumber \\
 \text{where}~~~\Psi(s,x)&=&s+\frac{n}{t}\ln \tilde{f}(s)= s + \frac{xD}{2 \pi^2} \ln \left[\text{sech}\left( 2\pi\sqrt{\frac{s}{D}}\right) \right]
 ~~~\text{with}~~~x=\frac{2\pi^2}{D}\left( \frac{n}{t}\right),\label{Psi}
\end{eqnarray}
because the survival probability $q(\tau_{last})$ in the remaining last time interval in Eq. (\ref{JPDF1}) do not 
contribute at the leading order. 
Performing a saddle point calculation for large $t$, one can see that the distribution $P(n,t)$ has the following large deviation form
\begin{equation}
 P(n,t)  \approx  e^{-t\mathcal{G} \left(x\right)};~~\text{with}~~~x=\frac{2\pi^2n}{Dt}, \label{LDP_n}
\end{equation}
as mentioned in the introduction. 
The large deviation function $\mathcal{G}(x)$ can be obtained from the minimum of $-\Psi(s,x)$ for fixed $x$ \emph{i.e.} 
\begin{equation}
 \mathcal{G}(x)= -\Psi(s^*,x) = -s^* - \frac{xD}{4\pi^2} \ln \left(1-\frac{4\pi^2s^*}{Dx^2} \right), \label{Gx+}
\end{equation}
where $s^*$ is the solution of $\frac{\partial \Psi(s,x)}{\partial s}\big{|}_{s^*}=0$, which in turn implies 
\begin{equation}
 s^*=\frac{D\alpha^{*2}}{4\pi^2},~~~\text{with}~~~\frac{\alpha^*}{\tanh \alpha^*} =x. \label{alpha*}
\end{equation}
for given $x$. Here we observe that this transcendental equation has solution $\alpha^*$ only for $x\ge1$ which means $s^*$ 
obtained from this solution can provide $\mathcal{G}(x)$ for $x \ge 1$ only. This is because the Laplace transform $\tilde{P}(n,s)$ is defined for $s \ge 0$. 
To obtain $\mathcal{G}(x)$ for $x \in[0, 1]$ we need to analytically continue $\Psi(s,x)$ for negative $s$ and that is done by using 
$\text{sech}\left( 2\pi\sqrt{\frac{-|s|}{D}}\right)=\sec \left( 2\pi\sqrt{\frac{|s|}{D}}\right)$ in Eq. (\ref{Psi}).
Hence for $s<0$, we have
\begin{equation}
 \Psi(s,x)|_{s < 0} = s + \frac{xD}{2 \pi^2} \ln\left[\sec\left( 2\pi\sqrt{|s|/D}\right) \right], 
 ~~~\text{where}~~~x=\frac{2\pi^2n}{Dt}. \label{Psi-s-ngtv}
\end{equation}
Following the same calculation as done for $s \ge 0$, we get 
\begin{eqnarray}
 \mathcal{G}(x) &=& -\Psi(-|s|^*,x)= |s|^* - \frac{xD}{4\pi^2} \ln \left(1+\frac{4\pi^2|s|^*}{Dx^2} \right), \label{Gx-} 
\end{eqnarray}
where $|s|^*$ is obtained from 
\begin{equation}
 |s|^*=\frac{D\beta^{*2}}{4\pi^2},~~~\text{with}~~\frac{\beta^*}{\tan \beta ^*} =x, \label{beta*}
\end{equation}
for given $x$. So first solving Eq. (\ref{alpha*}) for $x \ge 1$ and Eq. (\ref{beta*}) for $x \le 1$ and then using these solutions in Eqs. (\ref{Gx+}) and (\ref{Gx-}), 
respectively, we get the large deviation function $\mathcal{G}(x)$ for $x \in [0,\infty]$. Solving Eqs. (\ref{alpha*}) and (\ref{beta*}) analytically 
for arbitrary values of $x$ is not easy. Instead we solve these transcendental equations numerically 
and use those solutions to compute $\mathcal{G}(x)$ from Eqs. (\ref{Gx+}) and (\ref{Gx-}). 
In Fig. (\ref{fig5}), we plot the large deviation function $\mathcal{G}(x)$ as a function of $x$. 

One can find the form of $\mathcal{G}(x)$ explicitly in the following three limits (i) $x \to 0$, (ii) $x \to 1$ and (iii) $x \to \infty$ as 
\begin{eqnarray}
 \mathcal{G}(x) \simeq 
 \label{Gx-asmtry}
 \begin{cases}
  & \frac{D}{16} + \frac{D}{2\pi^2}~x\ln x,~~~~~x \to 0, \\
  &\\
  & \frac{3D}{8\pi^2}(x-1)^2,~~~~~~~|x-1| \to 0, \\
  &\\
  & \frac{D}{4\pi^2}x^2, ~~~~~~~~~~~~~~~~x \to \infty.
 \end{cases}
\end{eqnarray}
\begin{itemize}
 \item 
As $x\to 0$, the solution of Eq. (\ref{beta*}) is $\beta^* \simeq \frac{\pi}{2} - \frac{2x}{\pi}$ upto the leading order in $x$. Using this solution  
in Eq. (\ref{Gx-}), we get $\mathcal{G}(x) \simeq \frac{D}{16} + \frac{D}{2\pi^2}x\ln x$. 
The value $\mathcal{G}(0)= \frac{D}{16}$ at $x=0$,   
implies $P(0,t) \approx e^{-Dt/16}$ which means, the probability that the particle has not made any complete turn around the circle in time $t$, decays 
exponentially for large $t$. In the first exit picture, this is exactly the probability that the particle, starting from the origin,
stays inside the box $[-2\pi,~2\pi]$ till time $t$ and this probability is given by the survival probability $q(t)$ introduced in Eq. (\ref{FPndSVl}). 
Taking an ILT of $\tilde{q}(s)$ in Eq. (\ref{qtilde}), one can show that $q(t)$ for 
large $t$ indeed decays as $\sim e^{-Dt/16}$. 

\item For $x \to \infty$, the solution of Eq. (\ref{alpha*}) is $\alpha^* \simeq x[1-e^{-2x} + o(e^{-2x}) ]$ which implies 
$s^* \simeq \frac{Dx^2}{4\pi2}(1-e^{-2x})^2$. Putting this value of $s^*$ in Eq. (\ref{Gx+}) we get $\mathcal{G}(x) \simeq \frac{Dx^2}{4\pi^2}$ for large $x$.

For $x=1$, both Eqs. (\ref{alpha*}) and (\ref{beta*}) have solutions $\alpha^*=\beta^*=0$. Hence expanding the left hand sides of both the 
Eqs. (\ref{alpha*}) and (\ref{beta*}) for small $\alpha^*$ and for small $\beta^*$ respectively, we get
\begin{eqnarray}
 \alpha^* &\simeq& \sqrt{3(x-1)}~~~\text{for}~~x \gtrsim 1, \nonumber \\
 \beta^* &\simeq& \sqrt{3(1-x)}~~~\text{for}~~x \lesssim 1.
\end{eqnarray}
\begin{figure}[t]
\includegraphics[scale=0.3]{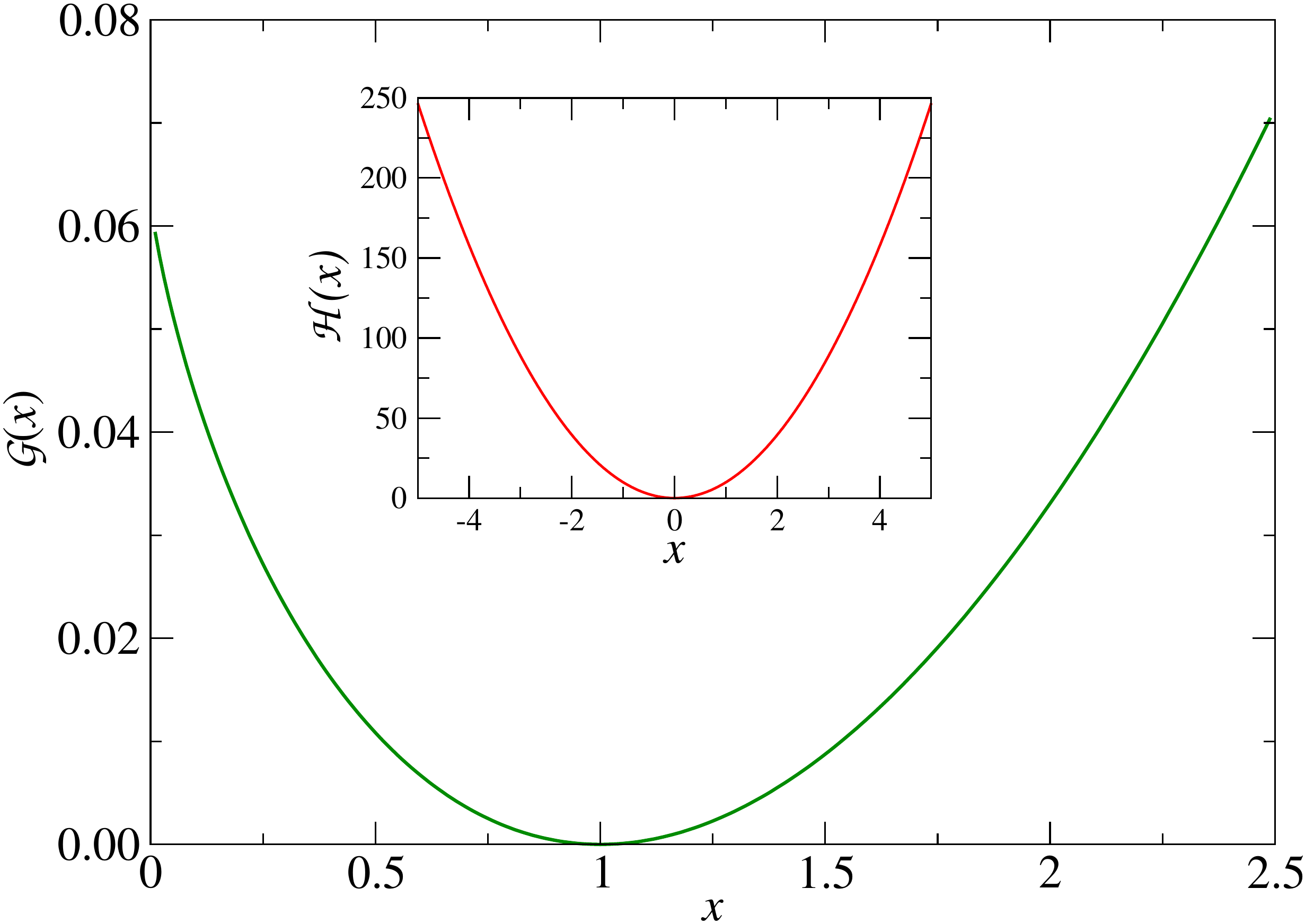}
\caption{(Color online) Plot of the large deviation function $\mathcal{G}(x)$ for $D=1$. Note the Gaussian nature of $\mathcal{G}(x)$ around $x \sim 1$ 
and the asymmetry in the shape between $x < 1$ and $x > 1$ as expressed in Eq. (\ref{Gx-asmtry}). 
Inset: Plot of the large deviation function $\mathcal{H}(x)$ for $D=1$.}
\label{fig5}
\end{figure}
Using these values of $\alpha^*$ and $\beta^*$ in Eqs. (\ref{Gx+}) and (\ref{Gx-}) respectively, 
one can show that the large deviation function $\mathcal{G}(x)$ in Eq. (\ref{LDP_n}) for $x \approx 1$ is given by $\mathcal{G}(x) \simeq \frac{3D}{8\pi^2}~(x-1)^2$
which implies 
\begin{equation}
 -\lim \limits_{t \to \infty} \frac{\ln P(n,t)}{t} = \frac{3\pi^2}{2D}\left(\frac{n}{t}-\frac{D}{2\pi^2} \right)^2,~~~~\text{for}~~n\approx \frac{Dt}{2\pi^2}.
\end{equation}
This large deviation form of $P(n,t)$ for $n$ around its mean $\frac{Dt}{2\pi^2}$, is consistent with the Gaussian scaling distribution $G(y)$ given in 
Eq. (\ref{Gauss}). 
\end{itemize}

We now put our attention on finding the large deviation form of the distribution $R(k,t)$ of the net winding number $k$. As done for $P(n,t)$, we start with 
\begin{equation}
R(k,t) =  \frac{1}{2\pi i} \int_{\mathcal{B}} ds~e^{st}~\tilde{R}(k,s) 
= \frac{1}{2\pi i} \int_{\mathcal{B}} ds~e^{st}~\frac{1}{s} \tanh\left(\pi\sqrt{\frac{s}{D}}\right)~\text{exp}\left(-2\pi |k|\sqrt{\frac{s}{D}} \right),
\end{equation}
where we have used the explicit form of $\tilde{R}(k,s)$ from Eq. (\ref{LTR-k}). In the $k \to \infty$ and $t \to \infty $ limit keeping $\frac{k}{t}$ finite, we see that 
the dominant contribution in the large deviation function comes from 
\begin{equation}
 R(k,t)  \approx \frac{1}{2\pi i} \int_{\mathcal{B}} ds~e^{st}~\text{exp}\left(-2 \pi |k| \sqrt{\frac{s}{D}} \right),
\end{equation}
where the rest of the terms in the integrand contribute at $\mathcal{O}(\frac{1}{t})$. Performing the Bromwich integral in the above equation we get the following 
large deviation form of $R(k,t)$
\begin{equation}
  R(k,t)  \approx e^{-t\mathcal{H}(k/t)}~~~\text{where}~~~\mathcal{H}(x) = \frac{\pi^2}{D}x^2. \label{LDF-R-k-t}
\end{equation}
This quadratic form of the large deviation function implies that over full range of $k \in [-\infty,\infty]$ the distribution $R(k,t)$ 
has a Gaussian scaling form under the scaling $\frac{k}{\sqrt{t}}$, which is given by $H(y)$ in Eq. (\ref{Gauss}) 
(see also the discussion in Sec.\ref{MVS-dis}). 

\subsection{Derivation of the exact distributions for arbitrary $n$ and $k$ }
\label{exact}
\noindent
Till now we have discussed about the asymptotic forms of the distributions $P(n,t)$ and $R(k,t)$ describing either typical or atypically large fluctuations. It turns out 
that one can find explicit expressions of the distributions $P(n,t)$ and $R(k,t)$ for arbitrary $n$, $k$ and $t$. Such explicit expressions for any $n$ and $k$ may 
be useful to compare with simulations. In the following we present the derivation of such expressions. 

To find an exact expression of $P(n,t)$ for any $n$ we need to perform the inverse Laplace transform in Eq. (\ref{ILTP_n}) which with the help of  
Eq. (\ref{LTP_n}) can be rewritten explicitly as 
\begin{eqnarray}
 P(n,t) &=& Q\left(n,\frac{Dt}{4\pi^2}\right)-Q\left(n+1,\frac{Dt}{4\pi^2}\right),~~~~~\text{where}, \nonumber \\
 Q(n,\tau) &=& \frac{1}{2\pi i} \int_{\mathcal{B}} dq~e^{q\tau}~\frac{1}{q}~\frac{1}
 {\left[\cosh \sqrt{q}\right]^{n}} .\label{ILTP_n-1}
\end{eqnarray}
Using the following expansion  
\begin{equation}
 \left[\cosh \sqrt{q}\right]^{-n}= \frac{2^n~e^{-n\sqrt{q}}}{(1+e^{-2\sqrt{q}})^n} 
 = \frac{2^n}{\Gamma[n]}\sum \limits_{\ell=0}^{\infty} (-1)^{\ell}\frac{\Gamma[n+\ell]}{\Gamma[\ell+1]}~e^{-(2\ell+n)\sqrt{q}},
 \label{expnsn}
\end{equation} 
and the identity \cite{Laplace-Joel}
\begin{equation}
 \frac{1}{2\pi i} \int_{\mathcal{B}} dq~e^{qt}~\frac{e^{-a\sqrt{q}}}{q}=\text{erfc}\left(\frac{a}{\sqrt{4t}}\right),~~~a>0,
\end{equation}
it follows that
\begin{equation}
Q(n,\tau)=   \frac{2^n}{\Gamma[n]}\sum \limits_{\ell=0}^{\infty} (-1)^{\ell}\frac{\Gamma[n+\ell]}{\Gamma[\ell+1]}~\text{erfc}\left(\frac{2\ell+n}{\sqrt{4\tau}}\right),
\end{equation}
where $\Gamma[x]$ is the Gamma function. 
Hence the probability distribution of having $n$ total winding around the circle in time $t$ is explicitly given by 
\begin{equation}
 P(n,t)=\frac{2^n}{\Gamma[n]}\sum \limits_{\ell=0}^{\infty} (-1)^{\ell}\frac{\Gamma[n+\ell]}{\Gamma[\ell+1]}~\text{erfc}\left(\frac{\pi(2\ell+n)}{\sqrt{Dt}}\right)-
 \frac{2^{n+1}}{\Gamma[n+1]}\sum \limits_{\ell=0}^{\infty} (-1)^{\ell}\frac{\Gamma[n+\ell+1]}{\Gamma[\ell+1]}~\text{erfc}\left(\frac{\pi(2\ell+n+1)}{\sqrt{Dt}}\right).
 \label{P-n-t-exact}
\end{equation}

\begin{figure}[t]
\includegraphics[scale=0.3]{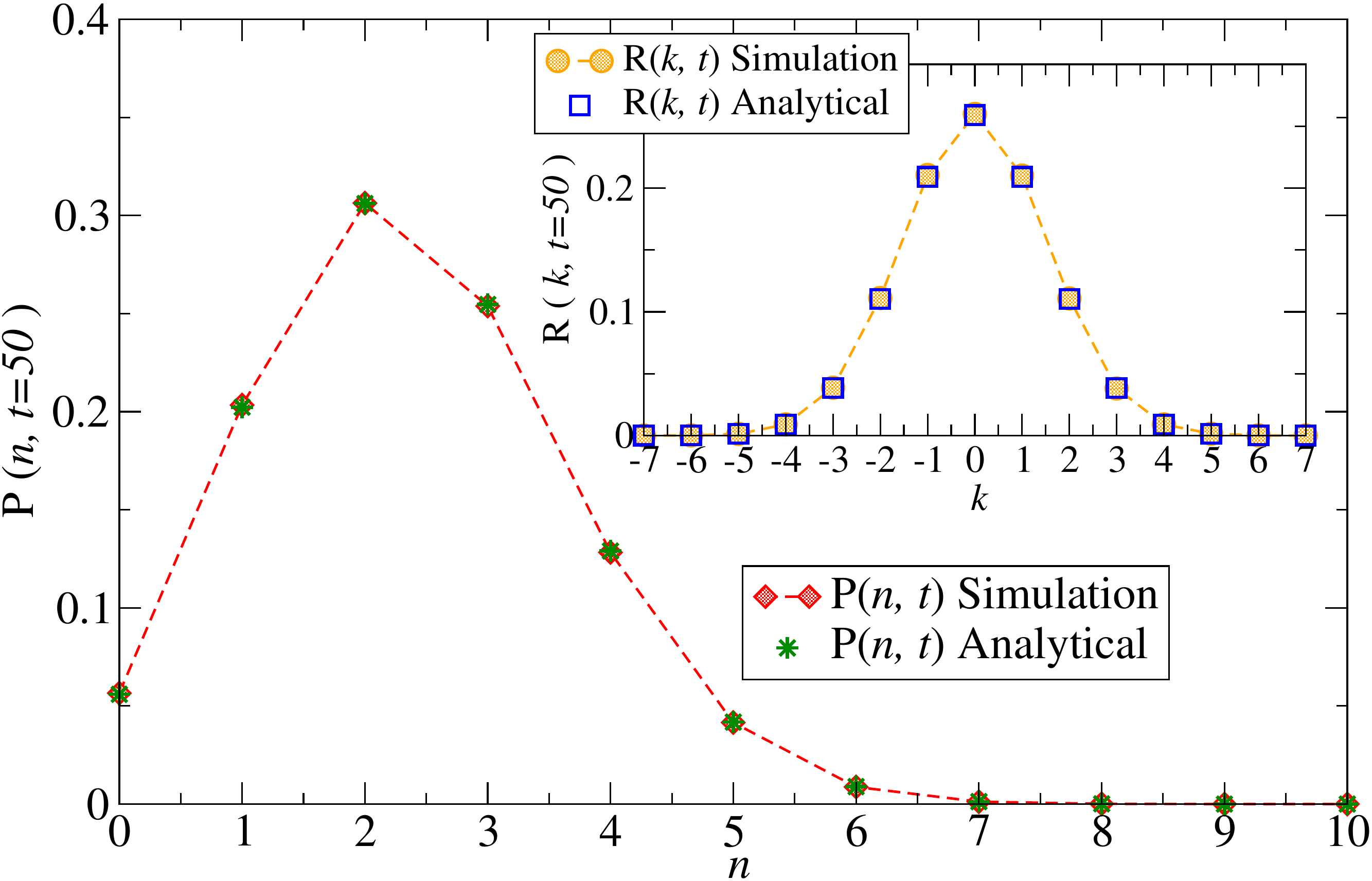}
\caption{(Color online) Comparison of the exact distributions $P(n,t)$ and $R(k,t)$ given in Eqs. (\ref{P-n-t-exact}) and (\ref{R-k-t-exact}), 
respectively, with the same obtained from numerical simulation of the Langevin Eq. (\ref{langevin}) for $t=50$ and $D=1$. Dashed lines are for guidance to the eye.}
\label{compare}
\end{figure}

Let us now turn our attention to the evaluation of $R(k,t)$ for arbitrary $k$. Once again this can be done by performing the inverse Laplace transform 
$R(k,t) = \frac{1}{2\pi i} \int_{\mathcal{B}} ds~e^{st}~\tilde{R}(k,s)$ where the explicit expression of $\tilde{R}(k,s)$ is given in Eq. (\ref{LTR-k}). Writing 
$\tilde{R}(k,s) = \tilde{A}_1(s)~\tilde{A}_2(s)$ where 
\begin{equation}
 \tilde{A}_1(s) = \frac{\text{exp}\left(-\frac{2\pi|k|}{\sqrt{D}}\sqrt{s}\right)}{\sqrt{s}},~~~\text{and}~~~
 \tilde{A}_2(s) =  \frac{1}{\sqrt{s}}~\frac{\sinh \left(\frac{\pi}{\sqrt{D}}\sqrt{s}\right)}{\cosh \left(\frac{\pi}{\sqrt{D}}\sqrt{s}\right)},
\end{equation}
we see that $R(k,t)$ can be expressed as a convolution $R(k,t) = \int_0^t d\tau A_1(\tau)A_2(t-\tau)$ of two functions $A_1(t)$ and $A_2(t)$ which are 
inverse Laplace transforms of $\tilde{A}_1(s)$ and $\tilde{A}_2(s)$ respectively, \emph{i.e.} $A_1(t) = \mathcal{L}_t^{-1}[\tilde{A}_1(s)]$ and  
$A_2(t) = \mathcal{L}_t^{-1}[\tilde{A}_2(s)]$. One can show that, $A_1(t)$ and $A_2(t)$ are explicitly given by 
\begin{equation}
 A_1(t)=\frac{\text{exp}\left(-\frac{\pi^2k^2}{Dt}\right)}{\sqrt{\pi t}},~~~\text{and}~~~
 A_2(t)=\frac{2\sqrt{D}}{\pi}\sum \limits_{m=0}^{\infty} \text{exp}\left( -\frac{(2m+1)^2Dt}{4}\right).
\end{equation}
Injecting these expressions in the convolution and performing the integral over $\tau$ we get 
\begin{equation}
R(k,t)= \frac{4}{\pi} \sum \limits_{m=0}^{\infty} \frac{(-1)^{k(2m+1)}}{(2m+1)}~\text{exp}\left( -\frac{(2m+1)^2Dt}{4}\right)~
\text{Im}\left[\text{erf}\left( \frac{\pi k}{\sqrt{Dt}}~+~i~\frac{(2m+1)\sqrt{Dt}}{2}\right) \right], \label{R-k-t-exact}
\end{equation}
where Im stands for imaginary part. Explicit expression of $\text{Im}\left[\text{erf}(x+iy)\right]$ is given
by \cite{Handbook}
\begin{equation}
 \text{Im}\left[\text{erf}(x+iy)\right] = \frac{2}{\sqrt{\pi}}e^{-x^2}\sum \limits_{l=0}^{\infty} \frac{(-1)^ly^{2l+1}}{(2l+1)!}~H_{2l}(x),
\end{equation}
where $H_m(x)$ is the $m$th order Hermite polynomial. Although this expression of $R(k,t)$ 
in Eq. (\ref{R-k-t-exact}) converges very fast numerically, it is not suitable for extracting the large $t$ Gaussian 
behavior described by the scaling function $H(y)$ in Eq. (\ref{Gauss}). However there is another representation 
$R(k,t)= \frac{2~e^{-\frac{\pi^2k^2}{Dt}}}{\pi \sqrt{\pi}} \sum \limits_{m=0}^{\infty} \frac{\mathcal{C}_{m,k}(t)}{(2m+1)}$ where 
 \begin{equation}
 \mathcal{C}_{m,k}(t) = \sum \limits_{\ell=0}^{\infty} (-1)^{\ell}~
  \frac{ 1.3.5...(2\ell-1)}{2^{\ell}\left(\frac{\pi^2 k^2}{Dt}~+~\frac{(2m+1)^2 Dt}{4}\right)^{2\ell +1}}~
  \text{Im}\left[\left(\frac{\pi k}{\sqrt{Dt}}~+~i~\frac{(2m+1)\sqrt{Dt}}{2}\right)^{2\ell +1} \right].
 \end{equation}
From this expression one can immediately see that in the large $k$ and $t$ limit keeping $k/t$ fixed, $R(k,t) \approx \frac{e^{-\frac{\pi^2k^2}{Dt}}}{\sqrt{t}}$ which is 
consistent with the scaling function $H(y)$ in Eq. (\ref{Gauss}) and also with the large deviation function $\mathcal{H}(x)$ in Eq. (\ref{LDF-R-k-t}). 
In Fig. \ref{compare} we compare the analytical expressions of the distributions $P(n,t)$ and $R(k,t)$ given in Eqs. (\ref{P-n-t-exact}) and (\ref{R-k-t-exact}), 
respectively, with the same obtained from numerical simulation for $t=50$ and $D=1$ and see very good agreement.

Another interesting quantity is the distribution of the maximum net winding number $k_{max}$ in time $t$. Denoting this distribution by $P_{max}(k,t)$, one 
can write $P_{max}(k,t)=Q_{max}(k,t)-Q_{max}(k-1,t)$ with $Q_{max}(-1,0)=0$, where $Q_{max}(m,t)=\text{Prob.}[k_{max} \le m,t]$. One can easily see that, $Q_{max}(m,t)$ 
is exactly the probability that the particle, starting at the origin, stays below the level $\theta=2\pi (m+1)$ throughout the time interval $[0,t]$. This means 
$Q_{max}(m,t)$ is the probability that the random process $\theta'(t)=2\pi (m+1)-\theta(t)$, starting from $\theta'(0)=2\pi (m+1)$, stays positive till time $t$ 
and it is given by the well known result \cite{Satya05,RednerBook, Bray13}
\begin{equation}
 Q_{max}(m,t)= \text{Erf}\left(\frac{2\pi (m+1)}{\sqrt{4Dt}}\right),
\end{equation}
using which we get  
\begin{equation}
 P_{max}(k,t)=Q_{max}(k,t)-Q_{max}(k-1,t) = \text{Erf}\left(\frac{\pi (k+1)}{\sqrt{Dt}}\right) - \text{Erf}\left(\frac{\pi k}{\sqrt{Dt}}\right)~~~\text{for}~~k=0,1,2...
\end{equation}

\section{Brownian Bridge on the ring}
\label{Brow-brdg}
In this section we consider the situation where the Brownian particle on the ring, starting from $\theta(0)=0$ \emph{i.e.} from the point A in Fig. \ref{fig1}, 
is constrained to come back to A after time $t$. This means that the final position of the particle is constrained to be 
$\theta(t)=2\pi l$ where $l$ is an integer. The probability $p_l(t)$ that the particle reaches $\theta(t)=2\pi l$ at time $t$ is given by 
$p_l(t)=\frac{\text{exp}\left(-\frac{\pi^2l^2}{Dt}\right)}{\sqrt{4\pi Dt}}.$
Given that such a constrained Brownian trajectory has net winding number $k$ in time $t$, implies that the final position of the particle at time $t$ 
is $\theta(t)=2\pi k$. 
Hence the probability distribution $R_c(k,t)$ of having net winding number $k$ (\emph{i.e} $k$ net counter-clockwise turns) in time $t$, 
is proportional to $p_k(t)$. Letting the normalization constant be $Z(t)$ one can write $R_c(k,t)= \frac{p_k(t)}{Z(t)}$ where 
\begin{equation}
 Z(t) = \sum \limits_{l=-\infty}^{\infty} p_l(t)=\frac{1}{\sqrt{4\pi Dt}} \sum \limits_{l=-\infty}^{\infty}\text{exp}\left(-\frac{\pi^2l^2}{Dt}\right)
 = \frac{1}{2\pi}\sum \limits_{m=-\infty}^{\infty} \text{exp}\left(-m^2Dt\right). \label{N_k-t}
\end{equation} 
From the expression of $R_c(k,t)= \frac{p_k(t)}{Z(t)}$ it is clear that this distribution is symmetric with respect to $k$ which implies, all odd order moments are zero. 
The lowest non-zero moment is $\langle k(t)^2 \rangle_c$ which can be computed and expressed in terms of $Z(t)$ in Eq. (\ref{N_k-t}) as
\begin{equation}
\langle k(t)^2 \rangle_c = \frac{Dt}{2\pi^2}+\frac{Dt^2}{\pi^2}\frac{d\ln Z(t)}{dt}. 
\end{equation}
It is easy to see that, the second moment of $k$ has the same large $t$ linear growth 
$\langle k(t)^2 \rangle_c \simeq \frac{Dt}{2\pi^2}$ as in the {\it free} case (see Sec.\ref{MVS-dis}). 

We now compute the probability $P_c(n,t)$ of having $n$ complete turns in time $t$ for Brownian bridges on the circle. As before (see Eq. (\ref{JPDF1}) )
let $\tau_1,~\tau_2,~\tau_3,....,\tau_n$ 
and $\tau_{\rm last}$ are time intervals in $t$ where $\tau_i$ represents the time required by the particle to perform the $i$th complete turn around the circle 
and $\tau_{\rm last}$ represents the time interval $\tau_{\rm last}=t-\sum_{i=1}^n\tau_i$ remaining after the $n$th complete turn. 
Following Sec. \ref{free} and 
remembering once again the connection between a complete turn and the first exit from the box of length $4\pi$, one can write 
\begin{equation}
P_c(n,t) = \frac{\mathcal{N}(n,t)}{Z(t)},~~~\text{with}~~~ 
\mathcal{N}(n,t)=\int_0^{t}\int_0^{t}...\int_0^{t} f(\tau_1)f(\tau_2)...f(\tau_n)~r_0(\tau_{last})~\delta \left( t-\tau_{\rm last}-\sum \limits_{i=1}^n \tau_i\right),
\label{mcalN-mult-t}
\end{equation}
where $f(t)$ is the first exit probability as before and $r_0(\tau_{\rm last})$ is the probability with which the particle starting from the origin 
(point A in Fig. \ref{fig1}) returns back to the origin in the last time interval $\tau_{\rm last}$ without making any complete turn. 
In the first exit picture, $r_0(\tau_{\rm last})$ represents the probability that the particle starting from the center of the box of length $4\pi$ (see Fig. \ref{fig1a}) 
comes back to the center in time $\tau_{\rm last}$ without exiting the box.
Taking Laplace transform of $\mathcal{N}(n,t)$ with respect to $t$ 
we get 
\begin{equation}
 \tilde{\mathcal{N}}(n,s) = \mathcal{L}_s[\mathcal{N}(n,t)] = \left[ \tilde{f}(s)\right]^n\tilde{r}_0(s), \label{LT-mcalN}
\end{equation}
where $\tilde{f}(s)= \mathcal{L}_s[f(t)]=\text{sech}\left (2\pi\sqrt{\frac{s}{D}} \right )$ from Eq. (\ref{ftilde}) 
and  $\tilde{r}_0(s)$ is the Laplace transform of $r_0(t)$ \emph{i.e.}
$ \tilde{r}_0(s) =\mathcal{L}_s[r(t)]$. 
To compute the probability $r_0(t)$, one needs to 
solve the diffusion equation 
\begin{equation}
\frac{\partial r(x,t)}{\partial t}=D \frac{\partial^2 r(x,t)}{\partial x^2}, \label{diff-r-xt}
\end{equation}
with BCs $r(x=\pm2\pi,t)=0$ and the initial condition $r(x,0)=\delta(x)$, where $r(x,t)$ represents 
the probability that the particle, starting at the origin, reaches $x$ at time $t$ while staying inside the box $[-2\pi,~2\pi]$ (see Fig. \ref{fig2}). 
Taking the Laplace transform with respect to $t$ on both sides of the above diffusion equation we get 
\begin{equation}
D \frac{\partial^2 \tilde{r}(x,s)}{\partial x^2} = s \tilde{r}(x,s) - \delta(x),~~~\text{with~~BC}~~ \tilde{r}(x=\pm2\pi,s)=0.
\end{equation}
It is easy to check that the solution of this equation is given by 
\begin{equation}
\tilde{r}(x,s) = \frac{1}{2\sqrt{s~D}} ~\frac{\sinh\left ((2\pi-|x|)\sqrt{\frac{s}{D}} \right )}{\cosh\left (2\pi \sqrt{\frac{s}{D}} \right )}, \label{LT-r_x} 
\end{equation}
which for $x=0$ provides $\tilde{r}_0(s) \equiv \tilde{r}(0,s)$. Once we know $\tilde{r}_0(s)$ and $\tilde{f}(s)$ explicitly, then taking inverse Laplace transform of 
$\tilde{\mathcal{N}}(n,s)$ in Eq. (\ref{LT-mcalN}) we get $\mathcal{N}(n,t)$ \emph{i.e.} 
\begin{equation}
\mathcal{N}(n,t)= \mathcal{L}_t^{-1}\left[ \left(\tilde{f}(s)\right)^n\tilde{r}_0(s)\right]=\frac{1}{2\pi i}\int_{\mathcal{B}}ds~e^{st}
~\left(\tilde{f}(s)\right)^n\tilde{r}_0(s)=\frac{1}{4\pi}~\frac{1}{2\pi i}\int_{\mathcal{B}}dq~e^{q\frac{Dt}{4\pi^2}}
~\frac{\sinh \sqrt{q}}{\sqrt{q}~[\cosh \sqrt{q}]^{n+1}}. \label{mcalN}
\end{equation}
At this point, one can easily check that 
$\sum_{n=0}^{\infty}\mathcal{N}(n,t) = Z(t)$ which implies $\sum_{n=0}^{\infty}P_c(n,t) = 1$. Moreover, using this expression of $\mathcal{N}(n,t)$ one can also  
find exact time dependence of the mean $\langle n(t) \rangle_c$ and the variance $\sigma^2_c(t)=\langle n(t)^2 \rangle_c- \langle n(t) \rangle^2_c$ (given  
in Appendix \ref{appndx-f}) from which we see that, for large $t$ both the mean and variance grow exactly 
the same linear way as in the {\it free} case (see Sec.\ref{MVS-dis}). 
This implies that for large $t$, the scaling distribution function corresponding to the distribution $P_c(n,t)$ is also described
by the same functions $G(y)$ as given in Eq. (\ref{Gauss}) for the {\it free} case. Furthermore, it can be easily seen that 
the large deviation function associated to $P_c(n,t)$ is same as the large deviation function $\mathcal{G}(x)$ 
associated to $P(n,t)$ (see Sec.\ref{LDFs}), because what happens in the last time interval $\tau_{last}$ after the $n$th
complete turn do not contribute at the leading order in the $n\to \infty$ and $t \to \infty$ limit while keeping $n/t$ fixed. 

To evaluate the inverse Laplace 
transform in Eq. (\ref{mcalN}), we follow the same steps as done in Sec.\ref{exact}. 
Once again using the expansion in Eq. (\ref{expnsn}) and the identity \cite{Laplace-Joel}
\begin{equation}
 \frac{1}{2\pi i} \int_{\mathcal{B}} dq~e^{qt}~\frac{e^{-a\sqrt{q}}}{\sqrt{q}}=\frac{1}{\sqrt{\pi t}}\text{exp}\left(-\frac{a^2}{4t}\right),~~~a>0,
\end{equation}
one can show that the function $\mathcal{N}(n,t)$ is explicitly given by 
\begin{equation}
\mathcal{N}(n,t) = \frac{1}{\sqrt{4\pi Dt}}~\frac{2^n}{\Gamma[n+1]}~\sum \limits_{\ell=0}^{\infty}(-1)^{\ell}~\frac{2\ell+n}{\ell+n}
~\frac{\Gamma[\ell+n+1]}{\Gamma[\ell+1]}~\text{exp}\left(-\frac{(2\ell+n)^2\pi^2}{Dt} \right),
\end{equation}
where $\Gamma[x]$ is the Gamma function. 
Hence we have an exact expression of the distribution $P(n,t)=\frac{\mathcal{N}(n,t)}{Z(t)}$ where $Z(t)$ is given in Eq. (\ref{N_k-t}).

Let us now look at the distribution $P_{max}^c(k,t)$ of the maximum net winding number $k_{max}$ in time $t$. Once again this distribution can be obtained from 
\begin{equation}
 P_{max}^c(k,t)=Q_{max}^c(k,t)-Q_{max}^c(k-1,t), \label{P_max}
\end{equation}
with $Q_{max}^c(-1,0)=0$ where $Q_{max}^c(m,t)=\text{Prob.}[k_{max} \le m,t]$. One can easily see that, the probability 
$Q_{max}^c(m,t)$ is equal to the ratio $Q_{max}^c(m,t)=\frac{\mathcal{N}_{max}^c(m,t)}{Z(t)}$. Here $\mathcal{N}_{max}^c(m,t)$ is the probability of having the particle's 
final position $\theta(t)$ at integer multiple of $2\pi$ while conditioned on the fact that the particle, starting from $\theta(0)=0$, 
stayed below the level $2\pi(m+1)$ throughout. The function 
$Z(t)$ in the denominator is given in Eq. (\ref{N_k-t}) and it represents the probability that the final 
position $\theta(t)$ of the particle is integer multiple of $2\pi$ given that the particle had started from $\theta(0)=0$. To evaluate $\mathcal{N}_{max}^c(m,t)$, 
we consider the shifted random process $\theta'(t)=2\pi(m+1)-\theta(t)$. The probability $\mathcal{N}_{max}^c(m,t)$ 
can be obtained from the propagator $g(\theta',t~|~2\pi(m+1),0)$ representing the probability density that the process $\theta'(t)$, 
starting from $\theta'(0)=2\pi(m+1)$, reaches $\theta'$ at time $t$ while staying positive throughout. This is actually the propagator of a Brownian particle 
with an absorbing wall at the origin and it is given by \cite{Satya05, RednerBook, Bray13} 
$g(y,t~|~x,0)=\frac{1}{\sqrt{4\pi Dt}} \left[\text{exp}\left( -\frac{(y - x)^2}{4Dt}\right)  - 
 \text{exp}\left( -\frac{(y + x)^2}{4Dt}\right)  \right]$.
Hence, the probability $\mathcal{N}_{max}^c(m,t)$ is given by 
$\mathcal{N}_{max}^c(m,t)=\sum_{\ell =1}^{\infty}g(2\pi \ell,t~|~2\pi(m+1),0)$. After performing some algebraic simplification 
of this infinite sum and using Eq. (\ref{N_k-t}), we get from Eq. (\ref{P_max}) :
\begin{equation}
P_{max}^c(k,t) = \frac{e^{-\frac{\pi^2k^2}{Dt}} +e^{-\frac{\pi^2(k+1)^2}{Dt} } }{\sum \limits_{l=-\infty}^{\infty}\text{exp}\left(-\frac{\pi^2l^2}{Dt}\right)}~~~
   \text{for}~~k\ge 0.
\end{equation}

\section{Conclusion}
\label{conclusion}
\noindent
Path integral techniques have been used to study statistics of net winding number $k(t)$ 
in many situations \cite{Edwards67, Edwards68, Comtet90, Nelson97} where one maps the problem to a suitable quantum problem.
But for other quantities like the total winding number $n$, path integral techniques are hard to adapt. 
In this paper we have presented a method based on renewal properties of Brownian motion to study winding statistics of a single Brownian motion on a ring.
This method is alternative to the standard path integral methods. 
More precisely, using the renewal property of Brownian motion and the connection between a complete turn around the circle and the first exit 
from a box of size $4\pi$, we derived analytical expressions of 
the probability distributions of the total number of turns $n$
and the net number of counter-clockwise turns $k$ at any time $t$.
Such distributions are relevant in quantum transport in ring geometry \cite{Texier}. 
For large $t$, we have shown that these distributions have Gaussian scaling forms describing the typical fluctuations of $\mathcal{O}(\sqrt{t})$ around their 
respective means. We have also computed the large deviation functions associated to these distributions, which describe the atypical fluctuations of 
$\mathcal{O}(t)$. Correlation between the total winding number $n$ and net winding number $k$ is studied from the joint probability 
distribution of $n$ and $k$ whose expression have been provided for any $t$. Numerical simulations have been performed to verify our analytical results.

One can extend this problem in different directions. For example, it would be interesting to see what happens to these distributions when the particle is being subjected 
to some pure or random potential. Investigating the equilibrium state of a ring coupled to a thermal bath reveals interesting connections with random walk 
on a Sinai potential \cite{Hurowitz}. In the context of polymer physics, fluctuations of winding number 
of a directed polymer in random media have been studied in \cite{Brunet}. 
Another interesting extension would be to consider interacting multiparticle system \cite{Wendelin}. Recently, winding statistics of $N$ non-intersecting 
Brownian bridges on unit circle have been studied for a case where the diffusion constant scales with the number of particles $N$ \cite{Litchy}. 
It will be interesting to study the winding statistics 
of non-intersecting walkers in the {\it{free}} case when the final positions of the walkers are not constrained. Finally, in the ring geometry, 
quantities other than the winding numbers, like residence time spent inside some given region, local time spent at some specified point etc. would also be interesting 
to study. Such quantities have been studied for general Gaussian stochastic processes in the context of persistence \cite{Bray13}.

This research was supported by ANR Grant No. 2011-BS04-013-01
WALKMAT and in part by the Indo-French Centre for
the Promotion of Advanced Research under Project
No.4604-3. We thank the Galileo Galilei Institute for Theoretical Physics, Firenze for the hospitality and support received.

\appendix
\section{Exact expressions of $\langle n(t) \rangle$, $\langle n(t)^2 \rangle$ and $\langle k(t)^2 \rangle$ for the free case}
\label{appndx}
Exact expressions of $\langle n(t) \rangle$, $\langle n(t)^2 \rangle$ and $\langle k(t)^2 \rangle$ are given as follows 
\begin{eqnarray}
 \langle n(t) \rangle &=&\langle k(t)^2 \rangle = \left(\frac{Dt}{2\pi^2}-\frac{1}{6}\right)+ \frac{D}{\pi^2} 
 \sum \limits_{m=1}^{\infty}\left(2+\frac{1}{Dm^2} \right) 
 e^{-Dm^2 t}=2 \sum \limits_{\ell=1}^{\infty}\ell~\text{erfc}\left(\frac{\pi\ell}{\sqrt{Dt}}\right), \label{mean-appndx} \allowdisplaybreaks[4] \\ 
 \langle n(t)^2 \rangle &=& \left(\frac{D^2t^2}{4\pi^4} + \frac{Dt}{6\pi^2} -\frac{2}{45}\right) + \frac{1}{2}\sum \limits_{m=1}^{\infty} 
  e^{-D m^2 t} \left(\frac{D t}{3 \pi ^2}+\frac{1}{6 \pi ^2 m^2}-\frac{D t}{2 \pi ^4 m^2}-\frac{1}{2 \pi ^4
   m^4}-\frac{2 D^3 m^2 t^3}{3 \pi ^4}\right) \label{second-appndx} \allowdisplaybreaks[1]\\
 &=& \frac{1}{3}\sum \limits_{\ell=0}^{\infty}~\frac{\Gamma[\ell+4]}{\Gamma[\ell+1]}~\left[\text{erfc}\left(\frac{(\ell+1)\pi}{\sqrt{Dt}}\right)
 + \text{erfc}\left(\frac{(\ell+2)\pi}{\sqrt{Dt}}\right) + \text{erfc}\left(\frac{(\ell+3)\pi}{\sqrt{Dt}}\right)\right].
\end{eqnarray}
Second expressions in both the above equations for $\langle n(t) \rangle$ and $\langle n(t)^2 \rangle $ are obtained using Poisson formula.

\section{Exact time dependence of the mean $\langle n(t) \rangle_c$ and $\langle n(t)^2 \rangle_c$ for the constrained case}
\label{appndx-f}

\begin{eqnarray}
&&\langle n(t) \rangle_c = 
\frac{\frac{Dt}{2\pi^2}+\frac{2Dt}{\pi^2}\sum \limits_{n=1}^{\infty}(1-2Dn^2t)e^{-Dtn^2}}{1+ 2\sum \limits_{n=1}^{\infty}~e^{-Dtn^2} }
=\sum \limits_{m=0}^{\infty}\frac{\Gamma[m+3]}{\Gamma[m+1]}
\frac{\left(e^{-\frac{(m+1)^2\pi^2}{Dt}} +e^{-\frac{(m+2)^2\pi^2}{Dt}}\right)}{\left(1+ 2\sum \limits_{l=0}^{\infty} e^{-\frac{l^2\pi^2}{Dt}}\right)}\nonumber \\
&&\langle n(t)^2 \rangle_c = \frac{\frac{D^2t^2}{4\pi^4}+\frac{Dt}{3\pi^2}
+\frac{4Dt}{\pi^2}\sum \limits_{n=1}^{\infty}(4D^3t^3~n^4-12D^2t^2~n^2-8\pi^2Dt~n^2+3Dt+4\pi^2)~e^{-Dtn^2}}{1+ 2\sum \limits_{n=1}^{\infty}~e^{-Dtn^2} }\nonumber \\
&&~~~~~~~~~~~= \frac{1}{12} \sum \limits_{m=0}^{\infty} \frac{\Gamma[m+5]}{\Gamma[m+1]}~
\frac{\left(e^{-\frac{(m+1)^2\pi^2}{Dt}} +3e^{-\frac{(m+2)^2\pi^2}{Dt}} +3e^{-\frac{(m+3)^2\pi^2}{Dt}}+e^{-\frac{(m+4)^2\pi^2}{Dt}}\right)}
{\left(1+ 2\sum \limits_{l=0}^{\infty} e^{-\frac{l^2\pi^2}{Dt}}\right)}.
\end{eqnarray}


\end{document}